\documentclass[twocolumn,showpacs,preprintnumbers,amsmath,amssymb,superscriptaddress,aps]{revtex4}

\usepackage{graphicx}
\usepackage{dcolumn}
\usepackage{bm}

\begin{document}


\title{Measurement of the Transverse Polarization of Electrons
\\ Emitted in Free Neutron Decay}

 \address{Marian Smoluchowski Institute of Physics, Jagiellonian University, Cracow, Poland}
 \address{Henryk Niewodnicza\'nski Institute of Nuclear Physics PAN, Cracow, Poland}
 \address{Paul Scherrer Institut, Villigen, Switzerland}
 \address{LPC-Caen, ENSICAEN, Universit\'e de Caen Basse-Normandie, CNRS/IN2P3-ENSI, Caen, France}
 \address{Katholieke Universiteit Leuven, Leuven, Belgium}

\author{A. Kozela}
 \address{Henryk Niewodnicza\'nski Institute of Nuclear Physics PAN, Cracow, Poland}

\author{G. Ban}
 \address{LPC-Caen, ENSICAEN, Universit\'e de Caen Basse-Normandie, CNRS/IN2P3-ENSI, Caen, France}
\author{A. Bia{\l}ek}
 \address{Henryk Niewodnicza\'nski Institute of Nuclear Physics PAN, Cracow, Poland}

\author{K. Bodek}
 \address{Marian Smoluchowski Institute of Physics, Jagiellonian University, Cracow, Poland}
\author{P.~Gorel}
 \address{Marian Smoluchowski Institute of Physics, Jagiellonian University, Cracow, Poland}
 \address{Paul Scherrer Institut, Villigen, Switzerland}
\address{LPC-Caen, ENSICAEN, Universit\'e de Caen Basse-Normandie, CNRS/IN2P3-ENSI, Caen, France}
\author{K.~Kirch}
 \address{Paul Scherrer Institut, Villigen, Switzerland}
\author{St.~Kistryn}
 \address{Marian Smoluchowski Institute of Physics, Jagiellonian University, Cracow, Poland}
\author{M.~Ku\'zniak}
 \address{Marian Smoluchowski Institute of Physics, Jagiellonian University, Cracow, Poland}
 \address{Paul Scherrer Institut, Villigen, Switzerland}
\author{O.~Naviliat-Cuncic}
 \address{LPC-Caen, ENSICAEN, Universit\'e de Caen Basse-Normandie, CNRS/IN2P3-ENSI, Caen, France}
\author{J.~Pulut}
 \address{Marian Smoluchowski Institute of Physics, Jagiellonian University, Cracow, Poland}
 \address{Paul Scherrer Institut, Villigen, Switzerland}
\author{N.~Severijns}
 \address{Katholieke Universiteit Leuven, Leuven, Belgium}
\author{E. Stephan}
 \address{Institute of Physics, University of Silesia, Katowice, Poland}
\author{J.~Zejma}
 \address{Marian Smoluchowski Institute of Physics, Jagiellonian University, Cracow, Poland}

\date{\today}

\begin{abstract}
Both components of the transverse polarization of electrons ($\sigma_{T_{1}}$, $\sigma_{T_{2}}$) 
emitted in the 
$\beta$-decay of polarized, free neutrons have been measured. 
The T-odd, P-odd correlation coefficient quantifying 
$\sigma_{T_{2}}$, perpendicular to the neutron polarization and 
electron momentum, was found to be $R=$ 0.008$\pm$0.015$\pm$0.005.  
This value is consistent with time reversal invariance, and significantly 
improves  limits on the relative strength of imaginary
 scalar couplings in the weak interaction.
The value obtained for the correlation coefficient associated with $\sigma_{T_{1}}$, $N=$ 0.056$\pm$0.011$\pm$0.005, agrees  with 
the Standard Model expectation, providing an important sensitivity test 
of the experimental setup.
\end{abstract}

\pacs{24.80.+y, 23.40.Bw, 24.70.+s, 11.30.Er}
\maketitle
Despite the great success of the Standard Model (SM) of elementary particles 
and their interactions, several important questions remain open.
One of these is the incomplete knowledge of physics of CP-violation, or via the CPT theorem, time reversal symmetry  violation (TRV).
The SM with the Cabbibo-Kobayashi-Maskawa (CKM) mixing scheme accounts for 
CP violation discovered in kaon \cite{Chris64} and B-meson 
\cite{Abe02,Aube02} systems.
It fails, however, to explain the basic observation of the 
dominance  of matter in the present Universe.
The discovery of new CP- or T-violating phenomena, especially in systems built 
of light quarks with vanishingly small contributions  
of CKM matrix induced mechanisms, would be a major breakthrough.
Nuclear beta decay experiments test these systems and free neutron 
decay plays a particular role: 
due to its simplicity it is free of corrections associated with 
the nuclear and atomic structure. 
Further, final state interaction effects, which can mimic T violation, are minimal and can, in addition, be calculated with a relative precision better than 1\% \cite{Vog83}. 

In this Letter we report on the first experiment searching for 
the real and imaginary parts of scalar and tensor  couplings via the  measurement of the transverse polarization of electrons emitted in the decay of free neutrons.
There exist very few measurements of this observable in general 
\cite{Dan05,Abe04}, and only two 
in nuclear beta decays. 
One of them, for the $^{8}$Li system \cite{Hub03}, 
provides the most stringent limit on tensor coupling constants of the weak 
interaction. 

According to \cite{jack57}, the decay rate distribution as a function of electron energy  ($E$)
and momentum (${\bf p}$), from polarized neutrons is proportional to: 
 \begin{equation}\label{Wprob}
  W({\bf J}, \mbox{\boldmath$ \hat{\sigma}$}, E, {\bf p}) \varpropto
 \dfrac{}{} 1  + \frac{\bf J}{J} \cdot
\left( A\frac{{\bf p}}{E} +R \frac{{\bf p} \times \mbox{\boldmath$ \hat{\sigma}$}}{E}+ N {\mbox{\boldmath$ \hat{\sigma}$}} \right) \dots
 \end{equation}
where ${\bf J}$ is the neutron spin, \mbox{\boldmath$\hat{\sigma}$} is 
a unit vector onto which the electron spin is projected and $A$ is beta decay asymmetry parameter. 
$N$ and $R$ are correlation coefficients which, for neutron decay 
with the SM assumptions  $C_{V}\!= \!C'_{V} \!= \!1$, 
$C_{A} \!= \!C'_{A}\! = \!\lambda \!=\! -1.27$, and allowing for a small admixture of scalar 
and tensor couplings $C_{S}$, $C_{T}$, $C'_{S}$, $C'_{T}$,  can be expressed as:
 \begin{eqnarray}\label{NR_N}
N &=& \!-0.218\cdot\! \Re(S) + 0.335\cdot\! \Re(T) - \frac{m}{E}\cdot\! A, \\
R &=& \!-0.218\cdot\! \Im(S) + 0.335\cdot\! \Im(T) - \frac{m}{137\, p}\cdot\! A,
\label{NR_R}
 \end{eqnarray}
where $S= (C_{S}+C'_{S})/C_{V}$, 
      $T= (C_{T}+C'_{T})/C_{A}$ and $m$ is the electron mass.
The $R$ correlation value vanishes to the lowest order within the SM.
Including final state interactions it becomes different from zero, $R_{FSI}\approx0.0006$, 
still below the sensitivity of the present experiment.  
A~larger measured value  would provide a hint for the existence of exotic couplings, 
and a new source of TRV. 

Applying the Mott polarimetry, both transverse  
components of the electron polarization can be measured simultaneously: 
$\sigma  _{T_{2}}$ perpendicular to the decay plane spanned by neutron spin and electron momentum, represented 
by $R$, and   $\sigma_{T_{1}}$ contained in the decay 
plane and associated with $N$.
The SM value of $N$ is finite and well within reach of 
this experiment. Its determination provides an important sensitivity test 
of the experimental apparatus.

The experiment was performed at the FUNSPIN beam line at the neutron source SINQ of the Paul 
Scherrer Institute, Villigen, Switzerland. 
The detailed description of the design, operation and performance of the 
Mott polarimeter can be found in \cite{Ban06}; here a short overview 
is presented.
The reported result comprises independent analyses of four data 
collection periods, featuring different basic conditions like beam 
polarization, Mott foil thickness and acquired statistics (see Table \ref{tab:table2}).

\begin{figure}
\includegraphics[scale=0.9]{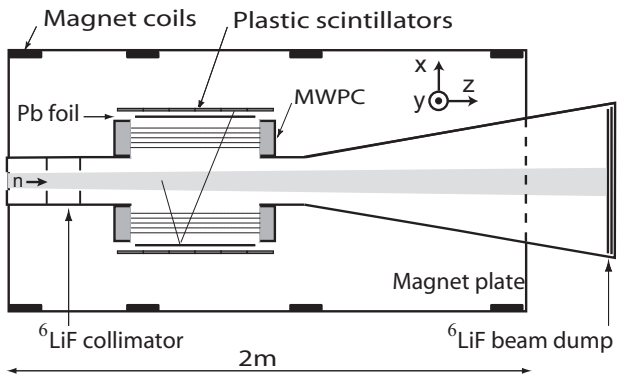}
\caption{\label{setup} Schematic top view of the experimental setup.  A~sample projection of an electron V-track event is indicated.}
\vspace*{-7pt}
\end{figure}

The Mott polarimeter consists of two identical modules, 
arranged symmetrically on both sides of the neutron beam (Fig.\,\ref{setup}). 
The whole structure was mounted inside a large volume dipole magnet 
providing a homogeneous vertical holding field of 0.5 mT within the beam
fiducial volume. 
An RF-spin flipper (not shown in Fig.\,\ref{setup}) was used to reverse the 
orientation of the neutron beam polarization at regular time intervals, typically every 16 s.
Going outwards from the beam, each module consists of a multi-wire proportional 
chamber (MWPC) for electron tracking, a removable Mott scatterer 
(1-2\,$\mu$m Pb layer evaporated on a 2.5 $\mu$m thick Mylar foil, almost transparent to electrons from neutron decay) and a 
scintillator hodoscope for electron energy measurement.

A 1 cm thick plastic scintillator allowed for the electron energy reconstruction with 33 keV FWHM resolution at 500 keV. 
The asymmetry of the light signal collected at both ends of the scintillator slab allowed for the determination of the vertical hit position 
with a resolution of about 6 cm, while the segmentation (10 cm) of the 
hodoscope in  horizontal direction provided a crude estimate of the z-coordinate.
Matching the information from the precise track reconstruction in the MWPC 
with that from the scintillator hodoscope considerably reduced background and 
random coincidences.

 A 1.3 m long multi-slit collimator defined the beam cross section to
4$\times$16 cm$^{2}$ at the entrance of the Mott polarimeter. 
In order to minimize neutron scattering and capture, the entire beam volume,
 from the collimator to the beam dump, 
was enclosed in a chamber lined with $^{6}$LiF polymer and filled with pure 
helium at atmospheric pressure. 
The total flux of the collimated beam was typically about 10$^{10}$ neutrons/sec. 
The beam divergence was 0.8$^{\circ}$ in horizontal and 
1.5$^{\circ}$ in vertical direction.
Thorough investigations of the beam polarization performed in a dedicated 
experiment \cite{Zej04} showed its substantial dependence on the  
position in the beam fiducial volume.
The average beam polarization necessary for the evaluation of the 
$N$- and $R$-correlation coefficients has been extracted from the observed 
decay asymmetry using the precisely known 
beta decay asymmetry parameter $A = -0.1173\pm0.0013$ \cite{PDG08}.
This approach automatically accounts for proper integration over the 
position dependent beam density, its polarization and detector acceptance.  
For this purpose single track events (only one reconstructed track segment on the hit scintillator side)
have been recorded using a dedicated prescaled trigger.
The main event trigger was used to find V-track candidates:
events with two reconstructed segments on one side and one segment accompanied by a
scintillator hit on the opposite side (Fig.\ \ref{setup}).

To extract the beam polarization $P$ the following asymmetries 
were analyzed:
 \begin{equation}\label{ASY1}
   \mathcal{E} \left(\beta, \gamma \right) = 
\frac
{N^{+}\left(\beta, \gamma \right) -N^{-}\left(\beta, \gamma \right)} 
{N^{+}\left(\beta, \gamma \right) +N^{-}\left(\beta, \gamma \right)}=
P \beta A cos(\gamma) ,
 \end{equation}
where $N^{\pm}$ are experimental, background corrected counts of single tracks, 
sorted in 4 bins of the electron velocity $\beta$ and 15 bins of the
electron emission angle  $\gamma$ with respect to the  neutron polarization direction.
The sign in superscripts reflects the beam polarization direction. 
Background counts, i.e. number of electrons not originating from neutron decay, 
were determined by comparing energy spectra of two event classes: (i) the reconstructed electron direction crossed the neutron 
beam (``from beam") and (ii) the electron origin was outside the neutron beam 
(``off beam''). 
The procedure relies on the assumption that the spectral shape of the background is 
the same for both event classes, while the characteristic neutron $\beta$-decay spectrum with 
end-point energy of 782 keV is 
present only in the ``from beam'' class.
For this assumption to hold, the ``off beam'' range has to be carefully chosen in both: 
inclination angle and extrapolated origin of the tracks on the opposite detector side \cite{Ban06}. 
The validity of this method was verified by comparing background-corrected energy spectra with simulated  $\beta$-decay spectra in which  energy loss and detector resolution were taken into account.
\begin{figure}[b]
\vspace*{-7pt}
\includegraphics[scale=.43]{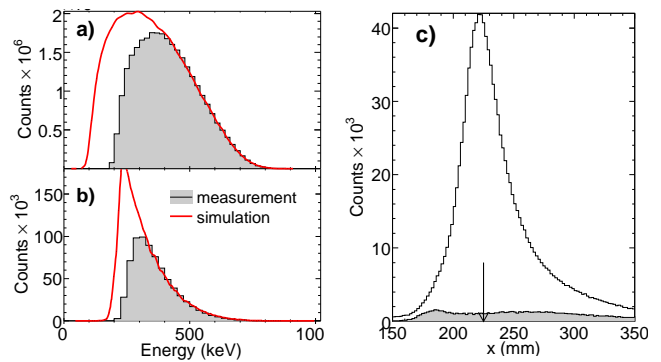}
\vspace*{-7pt}
\caption{\label{BetaSing} Background-corrected experimental energy 
	distributions (shaded areas) of (a) the single-track and (b) V-track events 
	compared with simulations. (c) Background contribution (shaded) to vertex $x$-coordinate 
	distribution of V-track events. The arrow indicates the Mott foil position.}
\vspace*{-5pt}
\end{figure}
Such a comparison is shown in Fig.\,\ref{BetaSing}a.
A similar background subtraction procedure was applied for the Mott scattering events (Fig.\,\ref{BetaSing}b). 
In the latter case 
the modification of the $\beta$ spectrum induced by the energy dependence of the Mott scattering cross section
is clearly visible.
Electronic thresholds are not included in the simulation --- this is why the measured and simulated distributions do not match at the low energy side.
The average neutron polarization values for the four data sets are collected  in Table \ref{tab:table2}.
The low polarization for the 2004 data set can be traced to a bug in the guiding field found {\em post factum} and verified in a dedicated experiment.

To extract the $N$ and $R$ correlation coefficients another set of 
asymmetries was considered:
 \begin{equation}\label{ASY3}
  \mathcal{A}\left(\alpha \right) = 
\frac
{n^{+}\left(\alpha \right) - n^{-}\left(\alpha \right)} 
{n^{+}\left(\alpha \right) + n^{-}\left(\alpha \right)} ,
 \end{equation}
where $n^{\pm}$ represent background-corrected experimental numbers of counts 
of V-track events, sorted in 12 bins of $\alpha$, defined as the angle between electron scattering and neutron decay planes. 
In the case of V-track events, beside the background discussed previously, 
events for which the scattering took place in the surrounding of the 
Mott-target provide an additional source of background. 
Fig.\,\ref{BetaSing}c presents the distribution of the reconstructed vertex 
positions in the $x$-direction for data collected with and without Mott-foil.
The distribution clearly peaks at the foil position.
The ``foil-out'' distribution has been scaled appropriately  by a factor 
deduced from the accumulated neutron beam.

It can be shown   \cite{Ban06} that: 
 \begin{equation}\label{ASY4}
   \mathcal{A}\left(\alpha \right) - P \bar{\beta} A  \mathcal{\bar{F}}(\alpha) = 
	 P \bar{S}(\alpha) \left[ N  \mathcal{\bar{G}}(\alpha) + 
	 R \bar{\beta}  \mathcal{\bar{H}}(\alpha) \right ] ,
 \end{equation}
where the kinematic factors $ \mathcal{\bar{F}}(\alpha)$, $ \mathcal{\bar{G}}(\alpha)$ and 
$ \mathcal{\bar{H}}(\alpha)$ represent the average values of the quantities 
${\bf \hat{J}\cdot\hat{p}}$,   ${\bf \hat{J}}\cdot\mbox{\boldmath$ \hat{\sigma}$}$ and
${\bf \hat{J}\cdot\hat{p}}\times\!\mbox{\boldmath$ \hat{\sigma}$}$, respectively,  $\bar{S}$ is the effective 
analyzing power of the electron Mott scattering, known in the literature as ``Sherman function'', and the bar over a letter indicates event-by-event 
averaging.
The term $P \bar{\beta} A  \mathcal{\bar{F}}$ 
accounts for 
the $\beta$-decay asymmetry induced nonuniform illumination of the Mott foil. 
Since the  $\bar{\beta}$ and $ \mathcal{\bar{F}}$ are known precisely from event-by-event averaging, the uncertainty of this term is dominated by the error 
of the average beam polarization $P$.  
To evaluate the influence of this term on the final result, the fit with the free parameters $R$ and $N$ was repeated with $P$ varied by one standard deviation. 
The obtained difference enters the budget of the systematic errors and is presented in 
Table \ref{tab:table1}.
\begin{table}[htb]
\vspace*{-12pt}
\caption{\label{tab:table1}Summary of systematic errors for the 2007 data set.}
\begin{ruledtabular}
\begin{tabular}{lccc}
Source &  $\delta N \times 10^{4}$  & $\delta R \times 10^{4}$ \\
\hline
term $P \bar{\beta} A  \mathcal{\bar{F}}$ &   5  &   23 \\
effective Sherman function $\bar{S}$      &  29  &   8  \\ 
guiding field misalignment                &  3   &   6  \\
background subtraction                    &  46  &   53 \\
dead time variations                      &   8  & 0.3  \\
\hline
Total                                     &  55  &   59 \\
\end{tabular}
\end{ruledtabular}
\end{table}

Mean values of the effective analyzing powers as a function of the electron 
energy as well as of scattering and incidence angles were calculated using the Geant 4 
simulation framework \cite{GEANT4}, following guidelines presented in \cite{Sal05,Kha01}. This approach allows accounting properly for atomic structure, nuclear size effects as well as for the effects introduced by multiple scattering in thick foils.  
The accuracy of these calculations has been  verified
\footnote{to be published elsewhere.} by comparison with two experimental 
data sets: at low (120 keV \cite{Ger91}) and high (14 MeV \cite{Srom99}) 
electron energies.

Mapping  of the spin holding magnetic field showed small 
nonuniformities in the beam fiducial volume. 
These were corrected for in the analysis. A residual systematic effect 
(see Table  \ref{tab:table1}) was induced by the uncertainty of the field measurements.

The systematic uncertainty is dominated by effects introduced by the background 
subtraction procedure, connected with the choice of the geometrical cuts defining event classes ``from-beam'' and ``off-beam''.
In order to estimate this effect, the cuts were varied in a range limited solely by the geometry of the apparatus.

Since the radio--frequency of the spin flipper was a source of small noise in the readout electronics, tiny spin flipper correlated dead time variations were observed. 
Their influence on the result was corrected for.
The residual effect is presented in Table  \ref{tab:table1}.

The asymmetries as defined in Eqs.\,\ref{ASY1} and \ref{ASY3} have been calculated for events 
with energy larger than the neutron $\beta$-decay end-point energy and for events originating outside the beam fiducial volume and were  found to be consistent with zero within statistical accuracy.  
This proves that the data were not biased by, for instance, a spin flipper related false asymmetry. 

\begin{figure}[hbt]
\vspace*{-8pt}
\includegraphics[scale=.43]{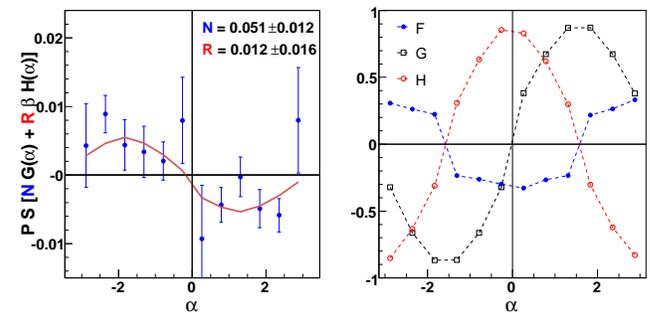}
\vspace*{-13pt}
\caption{\label{SNGRH}Left panel: experimental asymmetries $ \mathcal{A}$ corrected for the  $P \bar{\beta} A  \mathcal{\bar{F}}$ term for the 2007 data set as a function of $\alpha$ (defined in text). 
The solid line illustrates the two-parameter ($N$, $R$) 
least-square fit to the data. 
Indicated errors are of statistical nature.
Right panel: geometrical factors $ \bar{\mathcal{F}}(\alpha)$, $ \bar{\mathcal{G}}(\alpha)$ and $ \bar{\mathcal{H}}(\alpha)$ for the same data set. Dotted lines are to guide the eye only.}
\vspace*{-8pt}
\end{figure}

A fit of the experimental asymmetries $ \mathcal{A}$, corrected for the 
 $P \bar{\beta} A  \mathcal{\bar{F}}$ term, to the experimental data set of 2007 is 
shown in Fig.\,\ref{SNGRH}.
The  $R$ and $N$ coefficient values extracted in this way from all data sets are 
listed in Table \ref{tab:table2}.
\begin{table*}[htb]
\caption{\label{tab:table2}Summary of results obtained in all data collection periods. Statistical and systematic uncertainties follow the experimental values. 
$d$ is the nominal thickness of the Mott foil, $V_{n}$ represents the total, background corrected number of Mott-scattered events, $\bar{E}_{K}$ is the average  kinetic electron energy for those events and $N_{S\!M}$  is the SM value of the $N$ coefficient calculated at $\bar{E}_{K}$. Its error comes from the experimental uncertainty of the decay asymmetry parameter $A$ \cite{PDG08}.}
\begin{ruledtabular}
\begin{tabular}{ccccccccc}
Run  & $d$ ($\mu$m)  &  $V_{n}$   & $P\!\times\!10^{2}$    & $\bar{E}_{K}$(keV) & $N_{S\!M}\!\!\times\!\!10^{3}$ & $N\!\!\times\!\!10^{3}$(Eq.\,\ref{SR1})  &  $ N\!\!\times\!\!10^{3}$ (Eq.\,\ref{ASY4})  & $R \times\!10^{3}$ \\
\hline
2003 & 1             &  19000       & 80.3$\pm$1.3$\pm$1.6 & 331$\pm$1.0 $\pm$15  & 71$\pm$1 & 110$\pm$108$\pm$27         & 82$\pm$97$\pm$31     &  -89$\pm$143$\pm$38 \\
2004 & 1             &  74000       & 44.2$\pm$0.4$\pm$1.5 & 368$\pm$0.5 $\pm$15  & 68$\pm$1 & 144$\pm$92$\pm$15         & 70$\pm$86$\pm$17     &  -117$\pm$140$\pm$26 \\
2006 & 2             &  312000      & 80.0$\pm$1.0$\pm$1.5 & 365$\pm$0.2 $\pm$10  & 68$\pm$1 & 79$\pm$32$\pm$7         & 86$\pm$30$\pm$8     &  -11$\pm$42$\pm$9 \\
2007 & 2             &  1747000     & 77.4$\pm$0.2$\pm$0.7 & 370$\pm$0.1 $\pm$10  & 68$\pm$1 & 54$\pm$12$\pm$5          & 51$\pm$12$\pm$6      &  12$\pm$16$\pm$6 \\
\hline
Total &&2152000&&&&  59$\pm$11$\pm$4 &  56$\pm$11$\pm$5 &  8$\pm$15$\pm$5 \\
\end{tabular}
\end{ruledtabular}
\vspace*{-8pt}
\end{table*}

From the approximate symmetry of the detector with respect to the transformation 
$\alpha \rightarrow - \alpha$ it follows that  $ \bar{\beta}$, $\bar{S}$ and 
the factors $ \mathcal{\bar{F}}$, $ \mathcal{\bar{H}}$ are all symmetric 
while $ \mathcal{\bar{G}}$ is antisymmetric function of $\alpha$ (see 
Fig.\,\ref{SNGRH}). 
This allows the extraction of the  $N$ coefficient from the expression \cite{Ban06}:
\begin{equation}\label{SR1}
 N \approx  \frac{(r\!- \!1)}{(r\! +\!1)}\cdot
\frac{1-\frac{1}{2}(P\bar{\beta}A \bar{F})^2}{P \bar{S} \mathcal{\bar{G}}},\,\,\,
    r = \sqrt{\frac{n^{+}(\alpha) \, n^{-}(-\alpha )} 
		{n^{-}(\alpha )\,  n^{+}(-\alpha )} }
\end{equation}
The advantage of this method is that the effect associated with the 
term $P \bar{\beta} A  \mathcal{\bar{F}}$ is suppressed by a factor of about 60 as compared to Eq.\,\ref{ASY4}. 
The good agreement between the $N$ values obtained in both ways enhances 
confidence in the extracted $N$ and $R$ coefficient values.

\begin{figure}[hbt]
\includegraphics[scale=.54]{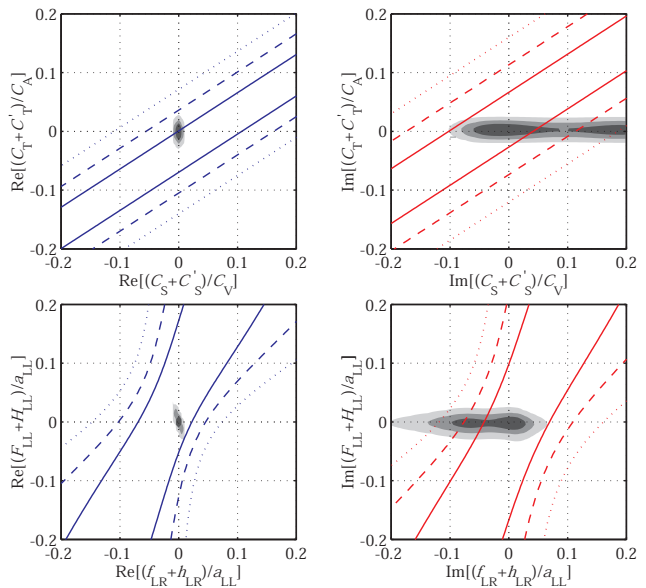}
\caption{\label{KB_rezultaty}Experimental bounds on the scalar vs. tensor normalized couplings (upper) and leptoquark 
exchange helicity projection amplitudes (lower panels). 
The grey areas represent the information as defined in
Ref. \cite{Sever06}, while the lines represent the limits resulting from the present experiment. Solid, dashed and dotted lines correspond to 1-, 2- and 3- sigma confidence levels, respectively, in analogy to decreasing intensity of the grey areas.}
\vspace*{-14pt}
\end{figure}

In Fig.\,\ref{KB_rezultaty} the new results have been included in 
exclusion plots containing all experimental information available 
from nuclear and neutron beta decays as surveyed in Ref. 
\cite{Sever06}. The upper plots contain the normalized scalar 
and tensor coupling constants $S$ and $T$ (see Eq.\,\ref{NR_N}), while 
the lower ones correspond to the helicity projection amplitudes 
in the leptoquark exchange model, as defined in Ref. \cite{Herc01}. 
This is the first determination of the $N$ correlation coefficient.
Although the present accuracy does not improve the already strong constraints on the real part of the couplings (left panels), the obtained result is consistent
with the existing data and, in addition, adds confidence to the validity of the extraction of $R$. 
As to the imaginary part (right panels), the new experimental value of the $R$ 
coefficient significantly constrains scalar couplings beyond the limits from all previous measurements.
The result is consistent with the SM.

This work was supported in part by Polish 
Committee for Scientific Research
under the Grant No.\,2P03B11122 and by an Integrated Action Program Polonium (Contract No.\,05843UJ).
Part of the computation work was performed at ACK Cyfronet, Krak\'ow.  
The collaboration is grateful to PSI for excellent support and kind hospitality.

\newpage 
\bibliography{PRL_nTRV}

\end{document}